\documentclass[10pt,conference,english,twocolumn]{IEEEtran} %

\usepackage{babel}
\usepackage[babel]{microtype}
\usepackage[babel]{csquotes}

\usepackage[T1]{fontenc}
\usepackage[svgnames]{xcolor}
\usepackage{graphicx}
\usepackage{tikz}
\usepackage{pgfplots}
\usepackage{subfigure}

\usepackage{tabularx}
\usepackage{booktabs}

\usepackage{algorithm}
\usepackage{algorithmic}

\usepackage{amsmath}
\usepackage{amsfonts}
\usepackage{amssymb}
\usepackage{amsthm}
\usepackage{bm}
\usepackage{siunitx}
\usepackage{mathtools}
\usepackage{dsfont}

\usepackage[math]{blindtext}
\usepackage[backend=biber,url=false,style=ieee,bibstyle=ieee-proc, dashed=false,isbn=false,doi=false,eprint=true]{biblatex}
\bibliography{main.bib}
\renewcommand*{\bibfont}{\small}
\usepackage{titlesec}

\titlespacing{\section}{0pt}{*0.5}{*0.5}

\usepackage{hyperref}
\hypersetup{
	pdfauthor={Irshad A. Meer, Karl-Ludwig Besser, Cicek Cavdar, H. Vincent Poor},
	pdftitle={Learning Based Dynamic Cluster Reconfiguration for UAV Mobility Management with 3D Beamforming},
	bookmarksnumbered=true,
    bookmarksopen=true,
	colorlinks=true,
	allcolors=.,
	urlcolor=MidnightBlue,
}

\usepackage[acronyms,nomain,xindy,nowarn]{glossaries}
\makeglossaries
\newacronym{5g}{5G}{Fifth-Generation Wireless Systems}
\newacronym{ai}{AI}{artificial intelligence}
\newacronym{ann}{ANN}{Artificial Neural Network}
\newacronym{ask}{ASK}{amplitude-shift keying}
\newacronym[plural={AWGN}, \glsshortpluralkey={AWGN}]{awgn}{AWGN}{additive white Gaussian noise}
\newacronym{au}{AU}{aerial user}
\newacronym{ap}{AP}{access point}
\newacronym{bec}{BEC}{binary erasure channel}
\newacronym{ber}{BER}{bit error rate}
\newacronym{bler}{BLER}{block error rate}
\newacronym{bpsk}{BPSK}{binary phase-shift keying}
\newacronym{bs}{BS}{base station}
\newacronym{bbu}{BBU}{baseband unit}

\newacronym{cdf}{CDF}{cumulative distribution function}
\newacronym{cnn}{CNN}{convolutional neural network}
\newacronym{csi}{CSI}{channel state information}
\newacronym{csit}{CSI\babelhyphen{nobreak}T}{channel-state information at the transmitter}
\newacronym{csir}{CSI\babelhyphen{nobreak}R}{channel-state information at the receiver}
\newacronym{comp}{CoMP}{coordinated multi-point}
\newacronym{cran}{C-RAN}{cloud-radio access network}

\newacronym{dnn}{DNN}{deep neural network}
\newacronym{ecc}{ECC}{error-correcting code}
\newacronym{ecdf}{ECDF}{empirical distribution function}
\newacronym{ee}{EE}{energy efficiency}

\newacronym{ff}{FF}{feed-forward}
\newacronym{ffnn}{FF-NN}{feed-forward neural network}
\newacronym{fsk}{FSK}{frequency-shift keying}

\newacronym{gm}{GM}{Gaussian mixture}

\newacronym{iid}{i.i.d.}{independent and identically distributed}
\newacronym{il}{IL}{information leakage}
\newacronym{iot}{IoT}{internet of things}

\newacronym{lb}{LB}{lower bound}
\newacronym{los}{LoS}{line-of-sight}

\newacronym{mac}{MAC}{multiple access channel}
\newacronym{map}{MAP}{maximum a posteriori}
\newacronym{mc}{MC}{Monte Carlo}
\newacronym{mdp}{MDP}{Markov decision process}
\newacronym{mimo}{MIMO}{multiple-input multiple-output}
\newacronym{miso}{MISO}{multiple-input single-output}
\newacronym{ml}{ML}{machine learning}
\newacronym{mmwave}{mmWave}{millimeter wave}
\newacronym{mop}{MOP}{multi-objective programming problem}
\newacronym{mrc}{MRC}{maximum ratio combining}
\newacronym{msc}{MSC}{modulation and coding scheme}
\newacronym{mse}{MSE}{mean squared error}
\newacronym{msac}{MSAC}{masked soft actor-critic}

\newacronym{nlos}{NLoS}{non-line-of-sight}
\newacronym{nn}{NN}{neural network}
\newacronym{nve}{NVE}{normalized validation error}
\newacronym{ric}{Near-RT RIC}{Near-Real Time RAN Intelligent Controller}
\newacronym{noma}{NOMA}{Non-orthogonal multiple access}

\newacronym{oru}{O-RU}{O-RAN radio unit}
\newacronym{oran}{O-RAN}{open-radio access network}
\newacronym{odu}{O-DU}{O-RAN distributed unit}

\newacronym{pdf}{PDF}{probability density function}
\newacronym{pwc}{PWC}{polar wiretap code}

\newacronym{qam}{QAM}{quadrature amplitude modulation}
\newacronym{qos}{QoS}{quality of service}

\newacronym{ra}{RA}{rearrangement algorithm}
\newacronym{rbf}{RBF}{radial basis function}
\newacronym{relu}{ReLU}{rectified linear unit}
\newacronym{ris}{RIS}{reconfigurable intelligent surface}
\newacronym{rl}{RL}{reinforcement learning}
\newacronym{rnn}{RNN}{recurrent neural network}

\newacronym{sac}{SAC}{soft actor-critic}
\newacronym{sgd}{SGD}{stochastic gradient descent}
\newacronym{sgp}{SGP}{secure goodput}
\newacronym{simo}{SIMO}{single-input multiple-output}
\newacronym{sinr}{SINR}{signal-to-interference-plus-noise ratio}
\newacronym{siso}{SISO}{single-input single-output}
\newacronym{snr}{SNR}{signal-to-noise ratio}
\newacronym{svm}{SVM}{support vector machine}

\newacronym{tvd}{TVD}{total variation distance}

\newacronym{uav}{UAV}{unmanned aerial vehicle}
\newacronym{ub}{UB}{upper bound}
\newacronym{urllc}{URLLC}{ultra-reliable low latency communication}

\newacronym{v2v}{V2V}{vehicle-to-vehicle}
\newacronym{v2x}{V2X}{vehicle-to-everything}

\newacronym{zoc}{ZOC}{zero-outage capacity}
\setacronymstyle{long-short}
\glsdisablehyper

\definecolor{plot0}{HTML}{004488}
\definecolor{plot1}{HTML}{DDAA33}
\definecolor{plot2}{HTML}{BB5566}
\definecolor{plot3}{HTML}{000000}
\definecolor{plot4}{HTML}{AAAAAA}

\definecolor{emerald}{HTML}{06D6A0}
\DeclarePairedDelimiter{\abs}{\vert}{\vert}

\newcommand*{\e}{\ensuremath{\mathrm{e}}}

\newcommand*{\reals}{\ensuremath{\mathds{R}}}

\newcommand*{\Exp}{\ensuremath{\mathrm{Exp}}}

\newcommand*{\one}{\ensuremath{\mathds{1}}}

\let\vec\mathbf
\pgfplotsset{compat=newest}
\pgfplotscreateplotcyclelist{lineplot cycle}{ %
	{plot0, mark=*, thick, mark options=solid},
	{plot1, mark=triangle*, thick, mark options=solid},
	{plot2, mark=square*, thick, mark options=solid},
	{plot3, mark=diamond*, thick, mark options=solid},
	{plot4, mark=pentagon*, thick, mark options=solid},
}

\pgfplotsset{%
	betterplot/.style={
		width=.93\linewidth,
		height=.27\textheight,
		xlabel near ticks,
		ylabel near ticks,
		cycle list name=lineplot cycle,
		mark options=solid,
		xmajorgrids=true,
		xminorgrids=true,
		ymajorgrids=true,
		grid style={line width=.1pt, draw=gray!20},
		major grid style={line width=.25pt,draw=gray!30},
		legend cell align=left,
		legend style = {
			/tikz/every even column/.append style={column sep=0.33cm}
		},
	},
}
\usetikzlibrary{positioning,calc,external}
\DeclareSIUnit{\dBm}{dBm}
\newcommand{\todo}[2][]{\ignorespaces
	\if\relax\detokenize{#1}\relax
	{\color{red}[TODO: #2]}%
	\else
	{\color{red}[TODO (#1): #2]}%
	\fi
}

\definecolor{change}{HTML}{0096b8}

\theoremstyle{plain}%

\theoremstyle{definition}

\newtheorem*{prob*}{Problem Statement}

\theoremstyle{remark}

\newtheorem*{rem*}{Remark}

\newtheoremstyle{example}{\topsep}{\topsep}{}{}{\itshape}{.}{ }{}
\theoremstyle{example}

\newtheorem*{example*}{Example}

\addto\extrasenglish{%

}

\IfPackageLoadedTF{titlesec}{%
	\ifCLASSOPTIONconference%
	\titlespacing{\section}{0pt}{1.5ex plus 1.5ex minus 0.5ex}{0.7ex plus 1ex minus 0ex} %
	\titlespacing{\subsection}{0pt}{1.5ex plus 1.5ex minus 0.5ex}{0.7ex plus .5ex minus 0ex} %
	\else
	\titlespacing{\section}{0pt}{3.0ex plus 1.5ex minus 1.5ex}{0.7ex plus 1ex minus 0ex} %
	\titlespacing{\subsection}{0pt}{3.5ex plus 1.5ex minus 1.5ex}{0.7ex plus .5ex minus 0ex} %
	\fi
	
	\def\thesubsubsectiondis{\arabic{subsubsection})}
	\def\theparagraphdis{\alph{paragraph})}
	\titleformat{\subsubsection}[runin]{\normalfont\normalsize\itshape}{\thesubsubsectiondis}{.5em}{}[:]
	\titlespacing*{\subsubsection}{\parindent}{0ex plus 0.1ex minus 0.1ex}{1ex}
	\titleformat{\paragraph}[runin]{\normalfont\normalsize\itshape}{\theparagraphdis}{.5em}{}[:]
	\titlespacing*{\paragraph}{2\parindent}{0ex plus 0.1ex minus 0.1ex}{1ex}
}{}

\newcommand*{\power}{\ensuremath{P}}

\newcommand*{\reward}{\ensuremath{r}}

 \title{Learning Based Dynamic Cluster Reconfiguration for UAV Mobility Management with 3D Beamforming}
\author{%
	\IEEEauthorblockN{Irshad A. Meer\IEEEauthorrefmark{1}\IEEEauthorrefmark{2}, Karl-Ludwig Besser\IEEEauthorrefmark{1}, Mustafa~Ozger\IEEEauthorrefmark{2},
        Dominic~Schupke\IEEEauthorrefmark{3},
        H. Vincent~Poor\IEEEauthorrefmark{1},  Cicek Cavdar\IEEEauthorrefmark{2}}
	\IEEEauthorblockA{\IEEEauthorrefmark{1}Department of Electrical and Computer Engineering, Princeton University, USA}
	\IEEEauthorblockA{\IEEEauthorrefmark{2}Division of Communication Systems, KTH Royal Institute of Technology, Sweden}
 \IEEEauthorblockA{\IEEEauthorrefmark{3}
 Airbus Central Research and Technology, Munich, Germany}
	
	Email: \{iameer, karl.besser, poor\}@princeton.edu,
           dominic.schupke@airbus.com,
           \{ozger, cavdar\}@kth.se
}

\makeatletter
\def\thanksfootnote{\gdef\@thefnmark{}\@footnotetext}
\makeatother

\begin{document}
\maketitle
\begin{abstract}\noindent\boldmath
In modern cell-less wireless networks, mobility management is undergoing a significant transformation, transitioning from single-link handover management to a more adaptable multi-connectivity cluster reconfiguration approach, including often conflicting objectives like energy-efficient power allocation and satisfying varying reliability requirements.
In this work, we address the challenge of dynamic clustering and power allocation for unmanned aerial vehicle (UAV) communication in wireless interference networks.
Our objective encompasses meeting varying reliability demands, minimizing power consumption, and reducing the frequency of cluster reconfiguration.
To achieve these objectives, we introduce a novel approach based on reinforcement learning using a masked soft actor-critic algorithm, specifically tailored for dynamic clustering and power allocation.
\end{abstract}
\thanksfootnote{This work was supported in part by the CELTIC-NEXT Project, 6G for Connected Sky (6G-SKY), with funding received from Vinnova, Swedish Innovation Agency.
The work of K.-L. Besser is supported by the German Research Foundation (DFG) under grant BE\,8098/1-1.
The work of H. V. Poor is supported by the U.S National Science Foundation under Grants CNS-2128448 and ECCS-2335876.
}
\glsresetall

\section{Introduction}\label{sec:introduction}

In modern cell-less wireless network architectures, users are no longer tied to a single \gls{ap} but are served simultaneously in non-orthogonal multiple access scenarios by a large number of distributed \glspl{ap}~\cite{bassoy2017coordinated}.
This has led to a significant transformation of the traditional approach to mobility management from the conventional handover management towards a more dynamic cluster reconfiguration model~\cite{hu2022scalable}. 
Consequently, the conventional notion of coverage has evolved from being centered around individual cells to becoming user-centric.

In this new paradigm, users can now be seamlessly served by a cluster of multiple distributed \glspl{ap} using the same frequency-time resources. 
However, this cluster configuration must be adjusted dynamically for each user's mobility and stringent \gls{qos} requirements like reliability requirements. 
Additionally, the cluster configuration can have multiple, often conflicting, objectives at the same time.
One crucial aspect is to minimize total transmission power while ensuring the high reliability demanded by modern applications. 
Moreover, the reliability requirements are dynamic and dependent on the system's state.
For instance, in safety-critical scenarios, such as communication between a central controller and a \gls{uav} near an airport, the need for high reliability is paramount. 
The variation in reliability demands can also be linked to the dynamic nature of services over time, where each service necessitates a distinct level of reliability~\cite{Wang2023}.
However, this also implies that transmission power can be saved in states where the reliability requirements are not as strict.

Cooperation of \glspl{ap} to form clusters and serve the users can be achieved using diverse techniques, such as \gls{comp}~\cite{dai2021joint, Mei2019}, \gls{cran}~\cite{checko2014cloud}, and cell-free networks~\cite{bjornson2019making}.
However, achieving the optimal cluster scheme can lead to substantial computational overhead, with complexity growing exponentially with network size~\cite{bassoy2017coordinated}.
Additionally, devising an efficient power allocation scheme for these dynamic clusters in a wireless interference network poses an even more challenging task~\cite{Matthiesen2020powerallocation}.

We therefore need a different approach for dynamic cluster formation and optimal power allocation with variable reliability constraints. 
\Gls{ml}, particularly \gls{rl}, offers an attractive solution for such a dynamic problem. 
With the ability to learn from the environment, \gls{rl} can exploit particular properties of \gls{uav} communication networks.
This allows the agent to strategically leverage movement patterns and \gls{los}/\gls{nlos} channel conditions between the \gls{uav} and the \glspl{ap}.

In this work, we adopt \gls{oran} as the underlying wireless network architecture. 
The \gls{oran} architecture offers diverse network clustering possibilities, driving efficiency advancements.
By dividing the physical layer into \glspl{odu} and \glspl{oru} (same as \glspl{ap}), it enables advanced features like \gls{ml} capabilities \cite{garcia2021ran}.
Functional blocks like \gls{ric} support tasks such as \gls{qos} and radio connection management. 
Additionally, \gls{oran} facilitates network-wide control, promoting cooperation among both \glspl{oru} within the same \gls{odu} and across different \glspl{odu}.

In existing literature, dynamic clustering has been a subject of investigation.
In \cite{beerten2023cell}, a dynamic clustering using the channel gains between the user and the \glspl{oru} is performed in an \gls{oran} architecture.
While they do not consider learning, their approach is offset by an increased signal overhead and it requires a cluster reconfiguration for all users when adding or removing a user.
Beamforming vectors for dynamic clustering of \glspl{ap} for a terrestrial user are designed using \gls{rl} in \cite{al2020multiple}.
However, they do not consider aerial users with varying reliability constraints.
The work in \cite{Meer2023asilomar} considers varying reliability for aerial users, but restricts scenarios to non-interfering environments with a single user.

In this work, we propose a novel dynamic cluster configuration and power control scheme for a downlink \gls{uav} communication system with varying reliability requirements.
The main contributions are summarized below.

\begin{itemize}
    \item We introduce a problem to optimize dynamic clustering and \gls{ap} power allocation within a wireless interference network.
    This aims to simultaneously satisfy time-varying reliability demands, minimize power usage, and reduce the frequency of cluster reconfiguration.

    \item We propose to solve the problem of dynamic clustering and power allocation with the \gls{sac} framework.
    To accommodate the dynamic nature of the state and action space, we employ an action masking technique, enabling our scheme to seamlessly handle the addition or dropping of new users without disrupting already formed clusters.
 
    \item We study and compare different performance metrics of the conventional clustering mechanism under different parameters in numerical simulations.
\end{itemize}

\section{System Model and Problem Formulation}\label{sec:system-model}
\begin{figure}[t]
	\centering
	\begin{tikzpicture}
\begin{axis}[
    betterplot,
    height=.21\textheight,
    xmin=0,
    xmax=3,
    ymin=0,
    ymax=3,
    xlabel={$x$ Position [$\si{\km}$]},
    ylabel={$y$ Position [$\si{\km}$]},
    legend style={
                 at={(0.325,0.9)}, 
                font=\footnotesize,
            },
]

\addplot[only marks, mark=triangle*, mark size=2.5pt] table {
    x y
    0.4 0.6
    1.2 0.8
    2.1 1.3
    0.9 1.7
    2.3 2.0
    1.1 2.4
    2.0 0.5
    0.8 2.9
    0.6 0.4 
    0.8 1.2 
    1.3 2.1 
    1.7 0.9 
    2.0 2.3
    2.4 1.1 
    0.5 2.0 
    2.9 0.8 
};
\addlegendentry{Access Points};

\addplot[only marks, mark=*, very thick, draw=orange!80!black, fill=orange, mark size=2.5pt] table {
    x y  class
    0.8 0.8  uav1
    2.2 2.5  uav2
    2.6 1.2  uav3
};

\addplot[dashed, plot1, thick, fill=plot1, fill opacity=.2] coordinates {(1, 1) (2, 1) (2, 2) (1, 2) (1, 1)};
\node[plot1!95!black] (eps2) at (axis cs: 1.5, 1.5) {$\boldmath\varepsilon_{\textbf{max},2}$};
\node[plot1!95!black] at (axis cs: 1.5, 0.3) {$\boldmath\varepsilon_{\textbf{max},1}$};
\addplot[very thick, dashed, plot2] coordinates {(0.8, 0.8) (0.4, 0.6)};
\addplot[very thick, dashed, plot2] coordinates {(0.8, 0.8) (1.2, 0.8)};
\addplot[very thick, dashed, plot2] coordinates {(0.8, 0.8) (0.9, 1.7)};

\addplot[very thick, densely dotted, plot0] coordinates {(2.2, 2.5) (1.1, 2.4)};
\addplot[very thick, densely dotted, plot0] coordinates {(2.2, 2.5) (2.3, 2.0)};
\addplot[very thick, densely dotted, plot0] coordinates {(2.2, 2.5) (2.1, 1.3)};

\addplot[very thick, emerald] coordinates {(2.6, 1.2) (2.3, 2.0)};
\addplot[very thick, emerald] coordinates {(2.6, 1.2) (2.0, 0.5)};
\addplot[very thick, emerald] coordinates {(2.6, 1.2) (2.1, 1.3)};
\addplot[very thick, emerald] coordinates {(2.6, 1.2) (2.9, 0.8)};

\node[font=\small, shift={(0.1,-0.2)}] at (axis cs:0.8, 0.8) {UAV 1};
\node[font=\small, shift={(0.05,0.25)}] at (axis cs:2.2, 2.5) {UAV 2};
\node[font=\small, shift={(0.18,0.25)}] at (axis cs:2.6, 1.2) {UAV 3};
\end{axis}
\end{tikzpicture}
    \vspace*{-.7em}
	\caption{Considered communication scenario with fixed \glspl{ap} and moving \glspl{uav} at an altitude above the ground. Within the highlighted zone in the center, the reliability constraint is $\varepsilon_{\text{max},2}$, otherwise it is $\varepsilon_{\text{max},1}$.}
	\label{fig:uav-scenario}
\end{figure}
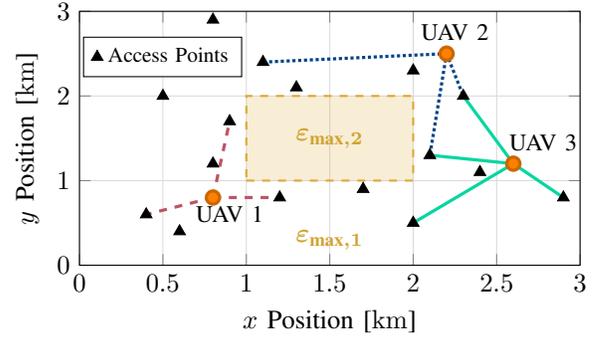

We consider an \gls{oran} architecture with a \gls{uav} downlink communication scenario, where communication is established from the \glspl{ap} to the \glspl{uav}, as depicted in \autoref{fig:uav-scenario}.
In a given area, we have $K$~\glspl{oru} (\glspl{ap}) deployed at fixed locations within a certain coverage area.
All the \glspl{oru} are connected to the O-Cloud with virtualization and processing resource-sharing capabilities \cite{demir2023cell}.
A total of $N$~\glspl{uav}, also referred to as \glspl{au}, are moving inside of the coverage area at the same time. 
All the \glspl{au} are equipped with a single antenna while each \gls{oru} is equipped with $L$~antennas.
When \gls{au}~$i$ enters the coverage area of the O-Cloud, a non-empty set of \glspl{oru}, $\mathcal{M}_{i}(t)\subseteq\{1, 2, \dots{}, K\}$, forms a cluster to serve \gls{au}~$i$.
Therefore, the total received power~$\power_i$ at \glspl{au}~$i$ at time~$t$ is given as
\begin{equation}
	\power_i(t) = \sum_{k=1}^{\abs{\mathcal{M}_i(t)}} h_{ik}(t) \, \power_{T,ik}(t) \, G\big(\theta_{i,k}(t), \phi_{i,k}(t)\big)\,, %
\end{equation}
where $\power_{T,ik}$ denotes the transmit power of \gls{oru}~$k$ to user~$i$, and $h_{ik}$ is the power attenuation between \gls{oru}~$k$ and user~$i$, i.e., the combined path loss and fading effects.
These effects are modeled according to~\cite{etsiLTEuav}.
With the known location of the user, we incorporate the 3D beamforming and beamtracking by leveraging the antenna radiation pattern and the steering vectors~\cite{Meer2023, colpaert20203d}.
For this, $G(\theta_{i,k}(t),\phi_{i,k}(t))$ represents the antenna array gain from \gls{oru}~$k$ to user~$i$, which is located at an elevation angle of $\theta_{i,k}(t)$ and azimuth angle of $\phi_{i,k}(t)$ with respect to the \gls{oru}.
The antenna array gain is given by
\begin{equation*}
G(\theta_{i,k}(t), \phi_{i,k}(t)) = G_{0} \cdot \vec{a}(\theta_{i,k}(t)) \cdot \vec{b}(\phi_{i,k}(t))\,,
\end{equation*}
where $G_{0}$ represents the constant array gain, while $\vec{a}(\theta_{i,k}(t))$ and $\vec{b}(\phi_{i,k}(t))$ represent the steering vectors in the elevation and azimuth directions, respectively.
The vectors $\vec{a}(\theta_{i,k}(t))$ and $\vec{b}(\phi_{i,k}(t))$ are given according to~\cite{colpaert20203d} as
\begin{align}
    \vec{a}\left(\theta_{i}\right) &= \sum_{m=1}^{M} \vec{I}_{m} \mathrm{e}^{\mathrm{j}(m-1)(k d_{z} \cos(\theta_{i}))}\\
   \vec{b}\left(\phi_{i}\right) &= \sum_{n=1}^{N} \vec{I}_{n}^{\sf{T}} \mathrm{e}^{\mathrm{j}(n-1)\left(k d_{y} \sin(\theta_{i}) \sin(\phi_{i})\right)}\,,
\end{align}
where, $\vec{I}_{m}$ and $\vec{I}_{n}$ denote column vectors of ones of sizes $m$ and $n$ respectively.
The number of antennas in $z$ and $y$ directions of the antenna array are $M$ and $N$, respectively.
The $d_{z}$ and  $d_{y}$ represent the antenna spacing in the $z$ and $y$ directions, respectively, and $k$ represents the wave number. 

For the ease of reading, we omit the time index~$t$, unless it is necessary to explicitly specify it.

With the above, the receive \gls{sinr} at the target \gls{au}~$i$ served by cluster $\mathcal{M}_{i}$ is:
\begin{equation}	
 \gamma^{\mathcal{M}_{i}}_{i} = \frac{\sum_{k=1}^{\abs{\mathcal{M}_{i}}} h_{ik} \power_{T,ik} G(\theta_{i,k},\phi_{i,k})}{N_{0} + \sum_{k=1}^{K} \sum_{\substack{n=1\\ n\neq i}}^{N}h_{ki} \power_{T,nk} G(\theta_{n,k},\phi_{n,k})},
 \label{eq:sinr}
\end{equation}
where $N_{0}$ is the noise density, and the interference power is the sum of all received power from all \glspl{oru} serving on the same resource to other \glspl{au}.

While we assume that the positions of the \glspl{au} and the fading statistics are known, the exact channel state is assumed unknown. 
Hence, the user will be in an outage with a non-zero probability when the \gls{sinr} at the \gls{au} is below a predefined threshold~$\gamma_{\text{th}}$, i.e., the outage probability for user~$i$ at time~$t$ is given as
\begin{equation}\label{eq:definition-outage-prob-sinr}
	\varepsilon_i(t) = \Pr\left(\gamma_{i}^{\mathcal{M}_i}(t) < \gamma_{\text{th}}\right).
\end{equation}
Depending on the specific use case, there exists an outage probability requirement, denoted as $\varepsilon_{\text{max}}$, that is deemed acceptable. 
However, it is essential to acknowledge that this tolerance level is influenced by various factors and may change over time, e.g., when the user moves into a different area. 
In this study, we focus on a particular scenario where a specific region is considered critical, demanding a higher level of reliability, denoted as $\varepsilon_{\text{max},2}$. 
Outside this critical area, it suffices for the outage probability to be lower than $\varepsilon_{\text{max},1}$, with $\varepsilon_{\text{max},1}>\varepsilon_{\text{max},2}$.

\subsection{SINR Outage}\label{sec:sinr-outage}
In the following, we derive an expression for calculating the outage probability from \eqref{eq:definition-outage-prob-sinr} for a single time slot~$t$, i.e., for a fixed power allocation and fixed positions of all users.
In this case, we can rewrite the outage probability as the probability of a new random variable that comprises of a sum of exponentially distributed random variables with different expected values,
\begin{align}
  \varepsilon_i(t) &= \Pr\left(\gamma_{i}^{\mathcal{M}_i}(t) < \gamma_{\text{th}}\right) \notag\\
           &=  \Pr\left(\sum_{k=1}^{\abs{\mathcal{M}_{i}}} h_{ik} \power_{T,ik} G(\theta_{i,k},\phi_{i,k}) < s_i\right)\notag \\
           &= \Pr\left(\sum_{k=1}^{\abs{\mathcal{M}_{i}}} Y_{ik} < s_i\right) \notag\\
           &= \Pr\left(T_i < s_i\right) \notag\\
           &= 1 - \bar{F}_{T_i}(s_i)
           \label{eq:sinr_outage}
\end{align}
where $s_i=\gamma_{\text{th}}\beta_i$ is the product of the \gls{sinr} threshold~$\gamma_{\text{th}}$ and the interference power~$\beta_i$ at user~$i$.
Based on the Rayleigh fading model, the random variable~$T_i$ is given as the sum of exponentially distributed variables~$Y_{ik}\sim\Exp(\alpha_{ik})$ with different expected values~$\alpha_{ik}$.
The expected values are given by the product of transmit power, antenna gain, and path loss.
The survival function~$\bar{F}_{T_i}$ of $T_i$ is given by~\cite{Amari1997}
\begin{align}
	\bar{F}_{T_i}(s) &= \sum_{k=1}^{K} A_{ik} \cdot \exp{(-\alpha_{ik}\cdot s)},\\
	A_{ik} &= \prod_{\substack{j=1\\j\neq k}}^{K} \frac{\alpha_{ik}}{\alpha_{ij}+\alpha_{ik}}, \quad \text{for } k=1, \dots, K.
\end{align}
For this expression to hold, we need to assume that all $\alpha_{ik}$ are distinct.
However, since they are the product of transmit power, antenna gain, and path loss, this assumption will hold almost surely in practice.

\subsection{Mobility Model}\label{sec:mobility-model}
In this work, we employ a realistic and tractable mobility model to capture the mobility of \glspl{uav}. 
In particular, we use the model provided in \cite{Smith2022} using coupled stochastic differential equations.
By utilizing estimated positions instead of actual ones, the model incorporates more realistic device trajectories and considers imperfect navigation. 
The advantage of this approach lies in its ability to generate smoother and realistic trajectories.
Additionally, it provides better control over velocity through correlation parameters, which influence stability and mobility based on distance-velocity relationships. 
Additional variance parameters scale Brownian perturbations, offering flexibility in introducing stochastic variations.
For detailed explanations of the model, please refer to the original work~\cite{Smith2022}.
Overall, this carefully chosen model accurately represents \gls{uav} mobility, accounting for practical considerations and enhancing the realism of trajectory generation.

\subsection{User Handling}
For a more realistic and dynamic scenario, our system needs to efficiently manage users entering and leaving the coverage area without disrupting the already established clusters.

\subsubsection{Users Entering the Coverage Area}
When a completely new mobile user enters the coverage area of the system, they were not previously associated with the system.
As a result, they become part of the active user group and are served by a new \gls{oru} cluster.
This ensures that new users seamlessly receive services within the existing system framework.

\subsubsection{User Leaves the Coverage Area}
On the other hand, active users can become inactive, e.g., by moving out of the service area or switching off their devices.
As a result, they transition from an active to a non-active state and are no longer served by the \gls{oru} cluster they were previously associated with.
This allows for efficient management of resources and ensures optimal service distribution to active users.

\subsection{Problem Formulation}\label{sub:problem-formulation} 
The seamless integration of dynamic clustering and power allocation for \glspl{oru} is of utmost importance to ensure both reliable and energy-efficient communication. 
To accomplish this objective, we present an optimization problem aimed at finding the optimal serving cluster for each user and the corresponding power allocation vector. 
The primary objective is twofold: The reliability of every user should be maximized, i.e., minimizing outages, while simultaneously minimizing the overall transmit power and the need for cluster reconfiguration arising from user movements. 
These two objectives are in conflict with each other since reducing the transmit power will in general lead to an increase of the outage probability.
Additionally, due to the movement of the users, optimal power allocation and the serving cluster vary over time.
Based on this, we use a scalarization to formulate the general multi-objective optimization problem of \glspl{oru} clustering and power allocation~as
\begin{subequations}
\label{eq:opt_prob}
\begin{align}
\label{eq:opt_prob-objective}
\min_{\power_{T,ik} , \mathcal{M}_{i}(t) } \; & \underbrace{Q_{1}(\mathcal{M}_{i}(t))}_\text{Cluster reconf.}
 + \underbrace{Q_{2}(\varepsilon_i(t))}_\text{Outage}
 + \underbrace{Q_{3}(\power_{T,ik})}_\text{Transmit Power} \\
\textrm{s.t.}\quad
& C_{1} : 0 \leq \power_{T,ik} \leq P_{\text{max}} \label{eq:constraint1}\\
& C_{2} : \abs{\mathcal{M}_{i}(t)} \geq 1 \label{eq:constraint2}
\end{align}
\end{subequations}
where $Q_{i}$, $i\in\{1, 2, 3\}$, are objective functions that measure the system's cost of cluster reconfiguration, reliability, and total transmit power, respectively.
The constraint $C_{1}$ states the transmitted power is limited by $P_{\text{max}}$, while $C_{2}$ makes sure all the users are served.
The exact choice of these functions will be described in the following \autoref{sec:learning-solution}.

The formulated optimization problem \eqref{eq:opt_prob} is non-convex.
Although employing sophisticated optimization techniques can potentially yield the globally optimal solution, the high complexity of such approaches poses significant practical challenges~\cite{Matthiesen2020powerallocation}.
This is particularly evident in scenarios characterized by dynamic channel variations, necessitating frequent updates to power control policies to ensure viable solutions.

\section{Proposed Solution With Reinforcement Learning}\label{sec:learning-solution}

In this section, we provide a \gls{rl}-based solution to solve problem \eqref{eq:opt_prob} by assuming that an agent is located at each \gls{ric} (also referred to as O-Cloud), which is connected with the \glspl{oru} in the coverage area.
The \gls{ric} with virtualization and processing resource sharing capabilities collects the required information from all the connected \glspl{oru}.
Using the trained model, it assigns a serving cluster with power allocation to all the \glspl{au}.  

In the context of interference networks, the action that the \gls{rl} agent takes, corresponds to a matrix of all transmit powers $\mathcal{A} \in \reals_{+}^{N \times K}$ for all \gls{oru}-user pairs.
The observation space consists of the current locations of all \glspl{uav}, the \gls{los}/\gls{nlos} conditions between each user and base station pair, and the status (active/inactive) of all the \glspl{uav}. 
Based on the action (power allocation) and observations (locations, \gls{los} condition), the outage probabilities~$\varepsilon_i$ for all users can be calculated.

We model our problem as a \gls{mdp} and transform the optimization problem as outlined in \eqref{eq:opt_prob} into the reward function within the \gls{rl} framework. 
We first describe the functions used in \eqref{eq:opt_prob-objective} to measure the system's cost of cluster reconfiguration, reliability, and transmit power.
The function that describes the system's cost of cluster reconfiguration~$Q_1$ is defined as:
\begin{equation}
     Q_{1}(\mathcal{M}_{i}(t)) = \frac{1}{N}\sum_{i=1}^{N}\one\Big( \mathcal{M}_i(t) \neq \mathcal{M}_i(t-1)\Big)
\end{equation}
which provides the proportion of clusters that have changed.
The function describing the reliability~$Q_2$ is defined as:
\begin{equation}
    Q_{2}(\varepsilon_i) = \frac{1}{N}\sum_{i=1}^{N}\one(\varepsilon_i > \varepsilon_{\text{max}}),
\end{equation}
which gives the fraction of users who are in the outage. 

Finally, the function that describes the system's transmit power~$Q_3$ is defined as:
\begin{equation}
    Q_3(\power_{T,ik}) = \frac{\sum_{i,k} \power_{T,ik}}{K \power_{T\text{max}}}, 
\end{equation}
which captures the total transmit power as a fraction of the maximum total transmit power.

To calculate the overall reward~$r$ and make it a positive quantity, we formulate the function as
\begin{multline}\label{eq:reward}
    \reward = \frac{\omega_1}{N}\sum_{i=1}^{N}\one\Big( \mathcal{M}_i(t) = \mathcal{M}_i(t-1)\Big)\\
 -  \frac{\omega_2}{N}\sum_{i=1}^{N}\one(\varepsilon_i > \varepsilon_{\text{max}}) 
 + \left(1 - \omega_3\frac{\sum_{i,k} \power_{T,ik}}{K \power_{T\text{max}}} \right)\,,
\end{multline}
where the non-negative weights $\omega_{i}$, $i\in\{1, 2, 3\}$, are used to balance between the individual objectives. 
The reward~$r$ in \eqref{eq:reward} increases when the stable clusters or non-outage users increase, while minimizing total transmit power.

As the learning algorithm, we employ the state-of-the-art \gls{rl} algorithm \gls{sac}, which optimizes the behavior of an agent given a state with the trial-and-error method. 
\Gls{sac} optimizes an agent's behavior through a trial-and-error approach, employing a \gls{dnn} policy that generates stochastic actions based on the state. 
It incorporates entropy regularization to promote exploration, balance exploration-exploitation trade-offs, and prevent premature convergence to suboptimal policies. 

To ensure efficient handling of dynamically changing observation and action space resulting from the movement of mobile users into and out of the service area, we employ a technique called \emph{action masking} \cite{huang2020closer}. Action masking involves setting the probability of allocating resources, such as power, to inactive users to zero. 
This effectively prevents the agent from taking actions that allocate resources to inactive users. Action masking is a valuable tool for streamlining the learning process, allowing the agent to focus on relevant actions while disregarding those that pertain to inactive users. 
In our context, this enables us to effectively manage the dynamic nature of users entering and leaving the coverage area. 
This targeted approach facilitates rapid adaptation of the agent's strategies and policies to the changing environment, leading to more effective resource management and improved overall performance.

We propose enhancing the \gls{sac} algorithm with action masking.
Throughout the following, we refer to this as \gls{msac}.
An overview can be found in \autoref{alg:masked_sac}.
This masking approach allows us to focus on relevant actions, excluding those associated with inactive users.
This optimization streamlines learning and empowers the agent to make more efficient decisions.
By combining the adaptability of \gls{sac} with strategic action masking, we effectively manage the environment's dynamic nature, accommodating varying user presence, and optimizing resource allocation for enhanced performance. However, we still need to fix a maximum number of users that can be supported in the coverage area.

\begin{algorithm}
\caption{Masked-SAC (MSAC) based Clustering}\label{alg:masked_sac}
\begin{algorithmic}[1]
\FOR{each episode $\leftarrow$ 1 to end}
\STATE Initialize: Observation space~$\mathcal{S}$ with location, \gls{los} condition and status of the $N$~\glspl{uav} 
\WHILE{not done}
\STATE $\mathcal{A}$: Action space from trained SAC agent 
\STATE Construct a mask $M$ based on the status of the users
\STATE Use that mask to get the final Action
$\mathcal{A'} = \mathcal{A} \cdot M$
\STATE Calculate the outage from \eqref{eq:sinr_outage} using $\mathcal{A'}$
\STATE Obtain the reward using \eqref{eq:reward}
\STATE Update status with new users or dropped users 
\ENDWHILE
\ENDFOR
\end{algorithmic}
\end{algorithm}

\section{Numerical Evaluation}\label{sec:simulations-per-eval}
\begin{table}%
\renewcommand*{\arraystretch}{1.2}
\centering
\caption{Hyperparameters employed for tuning our model}
\label{tab:hyper_parameters}
\footnotesize
\vspace{-5mm}
\begin{tabular}{p{0.2\textwidth} p{0.1\textwidth} p{0.1\textwidth}}\\
\toprule
\textbf{Parameters} & \textbf{Train} & \textbf{Test}\\ 
\midrule
Learning rate &    $10^{-5}$ & $10^{-5}$\\
Batch size &  \num{32768}  & \num{32768}  \\
Soft update coefficient $\tau$ &  $10^{-5}$ & $10^{-5}$ \\
Entropy coefficient  &  auto  &  auto \\
Iterations &  $2\cdot 10^6$  & $10^4$  \\
\bottomrule
\end{tabular}
\end{table}
In this section, we showcase the effectiveness of our proposed \gls{msac} implementation in solving \eqref{eq:opt_prob}.
Additionally, we benchmark our \texttt{MSAC} algorithm against two baseline methods. 
The first baseline is the \texttt{Opportunistic} cluster formation algorithm from \cite{beerten2023cell}, where \glspl{odu} opportunistically decides to include an \glspl{oru} in a cluster for user service based on channel gains. 
The second baseline follows the \texttt{Closest} strategy, where only the nearest \gls{oru} serves the user with maximum power. 
For a fair comparison, we keep all parameters of the communication system the same for all algorithms.
There are $N = 6$ users in a square area of {$\SI{3}{\km}\times \SI{3}{\km}$}.
They are served by $K = 19$ \glspl{oru}, which are placed randomly within this area with a height of $\SI{25}{\m}$, cf.~\autoref{fig:uav-scenario}. 
Each \gls{oru} is equipped with $L=16$ antennas.
The model of the path-loss follows \cite{etsiLTEuav}, where we set the carrier frequency to \SI{2.4}{\GHz} to accommodate a broader range for command and control traffic for the \glspl{uav}. 
Additionally, we set the \gls{sinr} threshold, denoted as $\gamma_{\text{th}}$, to \SI{-5}{\dB}.
The noise power at the receiver is given as $\sigma^2 = N_{0}B$, where $B = \SI{10}{\MHz}$ is the communication bandwidth, and ${N}_{0} = \SI{-174}{\dBm\per\Hz}$ is the noise spectral density.
The \glspl{uav} move randomly following the mobility model described in \autoref{sec:mobility-model} across the area.
The critical area with the higher reliability demand is located within {$[1, 2]\,\si{\km}$} in both $x$- and $y$-direction.
In this area, the outage requirement is set to {$\varepsilon_{\text{max},2}=10^{-5}$}, while it is {$\varepsilon_{\text{max},1}=10^{-2}$} everywhere else.

Leveraging the aforementioned parameters to generate channel information, we formulate our problem as a \gls{mdp} within OpenAI’s Gym environment framework. 
Following each iteration, the agent's policy generates values for the outage, transmit power, and cluster reconfiguration indicator.  
Using these values, the step reward \eqref{eq:reward} is calculated and fed back to the agent.  
The initial agent hyperparameters are summarized in \autoref{tab:hyper_parameters}, having been empirically determined through multiple iterations.
The source code of our implementation for reproducing all of the shown results is made publicly available at \cite{GithubCode}.

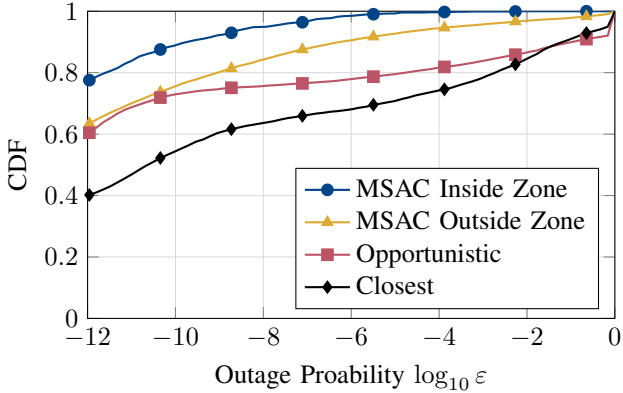
\begin{figure}[t]
    \centering
    \begin{tikzpicture}%
	\begin{axis}[
		betterplot,
		width=.97\linewidth,
		height=.24\textheight,
		xlabel={Outage Proability $\log_{10}\varepsilon$},
		ylabel={CDF},
		legend pos=south east,
		xmin=-12,
		xmax=0,
		ymin=0,
        ymax=1,
	]
          \addplot+[mark repeat=10] table[x=outage, y=inside, col sep=comma] {data/SAC_outage_data.csv};
    		\addlegendentry{MSAC Inside Zone};
		
		\addplot+[mark repeat=10] table[x=outage, y=outside,col sep=comma] {data/SAC_outage_data.csv};
		\addlegendentry{MSAC Outside Zone};
  
		\addplot+[mark repeat=10] table[x=outage, y=inside, col sep=comma] {data/Fixed_outage_data.csv};
		\addlegendentry{Opportunistic};

      \addplot+[mark repeat=10] table[x=outage, y=inside, col sep=comma] {data/Closest_outage_data.csv};
    		\addlegendentry{Closest};

	\end{axis}
\end{tikzpicture}
    \vspace*{-.5em}
    \caption{\Gls{cdf} of the outage probability~$\varepsilon$ experienced by the \glspl{uav}.}
    \label{fig:results-outage}
\end{figure}

For the described scenario above, the outage probability results can be found in \autoref{fig:results-outage}.
The \gls{cdf} shows the distribution of the outage probability for the users during their mobility in the service area. 
It can be observed that the proposed \gls{msac} algorithm outperforms the \texttt{Opportunistic} and \texttt{Closest} strategies. 
The proposed \gls{msac} performs better and learns about the stricter reliability constraint inside the high-reliability zone.
The agent can adopt the transmit power from the set of \glspl{oru} given the position of the users and \gls{los} conditions.
The proposed \texttt{MSAC} algorithm exploits the spatial relationships among the neighboring \glspl{oru} and \gls{los} conditions to mitigate interference and thus meet the requirements. 
It can be observed that inside the high reliability zone, we consistently meet the outage requirement of {$\varepsilon_{\text{max}, 2}=10^{-5}$} at all times.
Outside this zone, we achieve the $\varepsilon_{\text{max},1}=10^{-2}$ outage requirement approximately $\SI{98}{\percent}$ of the time. 
This behavior arises because the outage constraint is not strictly enforced in the reward function, but the agent can prioritize saving transmit power over meeting this requirement.
One way to address this challenge in future work is to change the reliability reward~$Q_{1}$ to employ a barrier-type function, e.g., a logarithmic function.
Although the \texttt{Opportunistic} outperforms the \texttt{Closest} strategy, the latter approach generates less interference by having only one \gls{oru} serve each user, but this impacts received power and subsequently increases the outage. 
However, both schemes are not able to differentiate between the high-reliability zone and meet the stricter requirements.

\begin{figure}[t]
    \centering
    \begin{tikzpicture}%
	\begin{axis}[
		betterplot,
		width=.97\linewidth,
		height=.24\textheight,
		xlabel={Fraction of the maximum transmit power},
		ylabel={CDF},
		legend pos=south east,
        legend style = {
            font=\small,
            fill opacity=.95,
            text opacity=1,
        },
		xmin=0,
		xmax=1,
		ymin=0,
        ymax=1,
	]

      \addplot+[mark repeat=10] table[x=X, y=cdf, col sep=comma] {data/power/SAC_power_data_in.csv};
    		\addlegendentry{MSAC Inside Zone};
       
        \addplot+[mark repeat=10] table[x=X, y=cdf, col sep=comma] {data/power/SAC_power_data_out.csv};
    		 \addlegendentry{MSAC Outside Zone};
		
		\addplot+[mark repeat=10] table[x=X, y=cdf, col sep=comma] {data/power/Fixed_power_data.csv};
    		\addlegendentry{Opportunistic};

	\end{axis}
\end{tikzpicture}
    \caption{Numerical results of the distribution of the total transmit power of the system normalized by the maximum available power.}
    \label{fig:results-ee}
\end{figure}
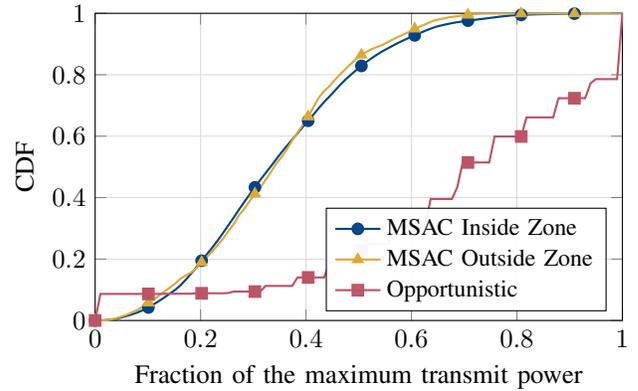

The results for the used power to compare the proposed \texttt{MSAC} with the \texttt{Opportunistic} are provided in \autoref{fig:results-ee}.
Since the transmit power is constant in the \texttt{Closest} scheme, we omit it from the graph.
To holistically assess the performance of \texttt{MSAC}, it is essential to consider the combined insights from both \autoref{fig:results-outage} and \autoref{fig:results-ee}.
The observations drawn from \autoref{fig:results-ee} indicate that the proposed \texttt{MSAC} utilizes less than $\SI{60}{\percent}$ of the maximum available transmit power in $\SI{90}{\percent}$ of the time. 
This underscores that the proposed scheme not only excels in mitigating outages and effectively distinguishes between varying reliability zones, but it also achieves these outcomes with a notable reduction in total transmitted power.

\begin{figure}
    \centering
    \begin{tikzpicture}
\begin{axis}[
    betterplot,
    width=.97\linewidth,
    height=.23\textheight,
    ylabel={PDF},
    xlabel={Number of \glspl{oru} per cluster},
    legend pos=north west,
]
\addplot+[ycomb, very thick] table [x=No, y=pdf, col sep=comma] {data/clusterSize/SAC_cluster_data.csv};
\addlegendentry{MSAC};

\addplot+[ycomb, very thick] table [x=No, y=pdf, col sep=comma] {data/clusterSize/Fixed_cluster_data.csv};
\addlegendentry{Opportunistic};
\end{axis}
\end{tikzpicture}
    \vspace*{-.5em}
    \caption{Average cluster size for serving mobile \glspl{uav}}
    \label{fig:result-clustersize}
\end{figure}
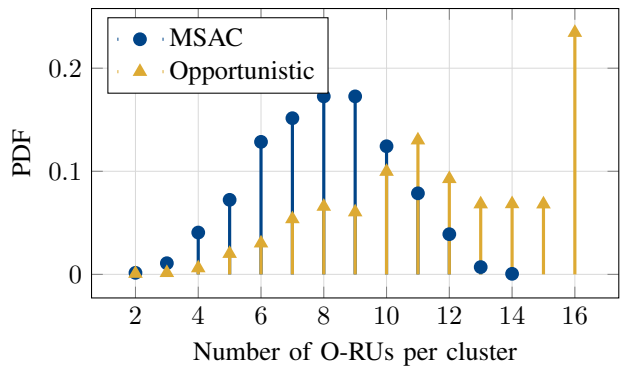

The third objective in our problem is the formation of clusters.
The distribution of the average cluster size for serving mobile \glspl{uav} is depicted in \autoref{fig:result-clustersize}.
Notably, the distribution for \texttt{MSAC} is centered, indicating its tendency to utilize half or fewer \glspl{oru} to form clusters.
Through training, the \texttt{MSAC} agent adeptly balances outage requirements and the number of \glspl{oru} per cluster.
In contrast, the \texttt{Opportunistic} scheme, driven by favorable channel conditions, often involves over $\SI{50}{\percent}$ of \glspl{oru} in clusters, resulting in excessive power usage.
Additionally, it is intriguing to note that \texttt{MSAC} maintains nearly zero probability for cluster sizes greater than or equal to $13$, whereas the \texttt{Opportunistic} scheme exhibits the highest probability for a cluster size of $16$, i.e., involving all \glspl{oru} in a cluster.

\section{Conclusion}\label{sec:conclusion}
Mobility management in the context of multi-connectivity involves real-time complex decision-making.
Dynamic cluster reconfiguration, compounded by time-varying reliability requirements and energy-efficient power allocation, presents a formidable challenge.

In this study, we have employed a model-free \gls{rl} algorithm to optimize dynamic cluster reconfiguration and associated power control under varying reliability demands.
Our primary objective has been to minimize the total transmit power of all \glspl{ap} within each cluster with minimum cluster reconfigurations while ensuring that outage probabilities remain below specified thresholds. 
These thresholds can change dynamically, such as when a \gls{uav} enters a critical zone with heightened reliability requirements.
Additionally, our approach accommodates the dynamic nature of observation and action spaces resulting from the arrival and departure of the mobile \glspl{au} to and from the service area. 
We have achieved this by enhancing the \gls{sac} algorithm through action masking. 

\printbibliography
\end{document}